\documentclass{iau}

\usepackage{aas_macros}
\usepackage{amsmath}
\usepackage[super]{nth}
\usepackage{graphicx}
\usepackage{multirow}
\usepackage[table]{xcolor}
\usepackage{xcolor}
\usepackage{orcidlink}
\usepackage{bbding}
\definecolor{black}{rgb}{0.0, 0.0, 0.0}\definecolor{red}{rgb}{0.0, 0.0, 0.0}\definecolor{green}{rgb}{0.0, 0.0, 0.0}\definecolor{blue}{rgb}{0.0, 0.0, 0.0}\definecolor{cyan}{rgb}{0.0, 0.0, 0.0}\definecolor{magenta}{rgb}{0.0, 0.0, 0.0}\definecolor{brown}{rgb}{0.0, 0.0, 0.0}\definecolor{gray}{rgb}{0.0, 0.0, 0.0}\definecolor{orange}{rgb}{0.0, 0.0, 0.0}\definecolor{pink}{rgb}{0.0, 0.0, 0.0}\definecolor{purple}{rgb}{0.0, 0.0, 0.0}\definecolor{teal}{rgb}{0.0, 0.0, 0.0}\definecolor{violet}{rgb}{0.0, 0.0, 0.0}\definecolor{olive}{rgb}{0.0, 0.0, 0.0}\definecolor{yellow}{rgb}{0.0, 0.0, 0.0}\definecolor{lime}{rgb}{0.0, 0.0, 0.0}

\begin{document}

\newcommand{\nbody}[1]{\textsc{Nbody#1}}
\newcommand{\dragon}[1]{\textsc{Dragon#1}}
\newcommand{\petar}{\textsc{PeTar }}
\newcommand{\bifrost}{\textsc{BiFrost }}
\newcommand{\co}{\mathcal{O}}
\newcommand{\tree}{\textsc{Tree}}

\lefttitle{Rainer Spurzem}
\righttitle{From NBODY1 to NBODY7}

\jnlPage{1}{7}
\jnlDoiYr{2021}
\doival{10.1017/xxxxx}

\aopheadtitle{Proceedings IAU Symposium}
\editors{H.M. Lee, J. Hong, R. Spurzem, eds.}

\title{From \nbody{1} to \nbody{7}: \\ the Growth of Sverre's Industry}

\author{Rainer Spurzem\orcidlink{0000-0003-2264-7203}}
\affiliation{Astronomisches Rechen-Institut, Zentrum f\"ur Astronomie, University of Heidelberg, M\"onchhofstr. 12-14, 69120 Heidelberg, Germany \\
National Astronomical Observatories, 
Chinese Academy of Sciences, 20A Datun Rd., Chaoyang District, 100012, Beijing, China \\
Kavli Institute for Astronomy and Astrophysics, Peking University, Yiheyuan Lu 5, Haidian Qu, 100871, Beijing, China}

\begin{abstract}
``From \nbody{1} to \nbody{6} : The Growth of an Industry'' is the title of a 1999 invited review by Sverre Aarseth, for Publications of the Astronomical Society of the Pacific (PASP). I took this as an inspiration for the title of this paper; it describes how Sverre's \nbody{} Industry has further grown since 90s of the previous century, and how it is further flourishing and hopefully developing, in his spirit, even after the sad news of his passing away reached us. My contact and friendship with Sverre started a few decades ago being sent to Cambridge to learn \nbody5, counting input parameters, and learning about the fact that even a sophisticated code (which had already at that time quite a history) requires permanent maintenance and bug fixes. Managed by Sverre, who relentlessly ran his code and responded to the widely spread crowd of "customer" colleagues. There has been a phase of massive and fast development and improvements due to vectorization, parallelization, GRAPE and GPU acceleration, and Sverre has been always on top of it if not ahead, but also fully adopting ideas of collaborators, once they tested well. 
\nbody{6++GPU} and \nbody{7} entered the scene, and also recent new competitors, such as \petar or \bifrost. We all have learnt a lot from Sverre, and strive to continue in his open-minded spirit, for open source and exchange. A striking evidence for the further growth of the ``industry'' is the number of papers here (and two of them follow in this session, but also in other sessions) using and further developing the aforementioned codes, as well as the occurrence of new and competing codes, which keep the field alive.
\end{abstract}

\maketitle
\section{Introduction}
This article has two sides. On one hand side there is the sad duty to provide a final and conclusive overview over Sverre Aarseth's ``\nbody{} industry'', on the other hand side there is the good news that his legacy continues to grow, the industry is alive. Here I want to present a personal and selective view about Sverre's work and life, some scientific achievements, and how I think the legacy continues at the current time and will continue into the future. The article concludes with a scientific summary of the current features of \nbody{6++GPU} (and to a lesser extent \nbody{7}, but see Banerjee, this volume), its performance and some exemplary science results. Also competing codes for direct and accurate \nbody{} simulations, which have been rare in the past decades, will be briefly quoted. 

\section{\nbody{} -- the beginning}
\label{beginning}
Sebastian von Hoerner found in his early published $N$-body simulations that the relaxation time \citep{Chandrasekhar1942} is relevant for star cluster evolution and that the formation of close and eccentric binaries occurs. It is particularly difficult to accurately integrate them, the simulation would practically stop, since close binaries demand too tiny time-steps \citep{vHoerner1960,vHoerner1963}.

About at the same time a young postdoc -- Sverre Aarseth -- in Cambridge developed a direct $N$-body integrator to simulate clusters of galaxies using a gravitational softening, thereby avoiding von Hoerner's problems with tight binaries and singularities \citep{Aarseth1963}. Regularization methods \citep{KustaanheimoStiefel1965} were later implemented in Aarseth's direct $N$-body code \citep{Aarseth1971}. This allowed to proceed past the binary deadlock detected in von Hoerner's models (see Sect.~\ref{regularization} below). Remarkably von Hoerner and Aarseth never met until their only meeting during the star2000 conference in Heidelberg \citep{Deitersetal2001}. During
that meeting they could for the first time compare their codes, and von Hoerner published a remarkable account on ``how it all started'' \citep{vHoerner2001}.

\nbody{5} \citep{Aarseth1985a} had already become an ``industry standard''. It used Taylor series using up to the third derivative of the gravitational force, in a divided difference scheme based on four time points, with individual particle time-steps (different from the current Hermite scheme, see below Sect.~\ref{Hermite}). Regularizations for more than two bodies used the classical chain regularization \citep{MikkolaAarseth1990}.

\begin{figure}
  \centering
  \includegraphics[width=0.7\textwidth]{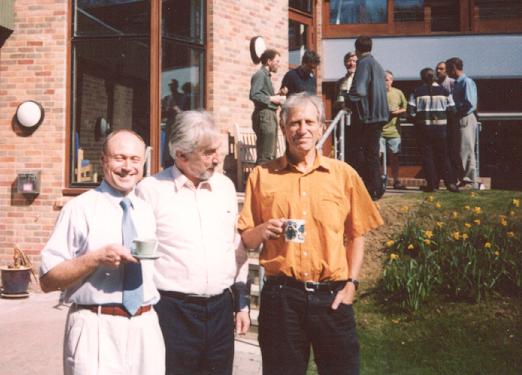}
 \caption{Sverre Aarseth (right) and Chris Tout (left), with visitor Emmanuil Vilkoviski of Fesenkov observatory in Almaty, Kazakhstan; in mid 90's.}
\label{fig:cambridge90s}
\end{figure}

Stellar evolution is a very successful ingredient to all \nbody{} code versions, pushed forward from early on by Jarrod Hurley, Chris Tout, and collaborators \cite[cf. e.g.][see also Hurley, this volume, and Fig.~\ref{fig:cambridge90s}]{Hurleyetal2000,Hurleyetal2001,Hurleyetal2002b,Hurleyetal2005}. While a historical account and the current state of the art is summarized here by Hurley (this volume), further details and new developments on massive stars, winds, and relativistic binaries can be found inside articles in \cite{Aarsethetal2008,Aarseth2003b} and in the recent review of \cite{SpurzemKamlah2023}. 

\begin{figure}
  \centering
  \includegraphics[width=0.6\textwidth]{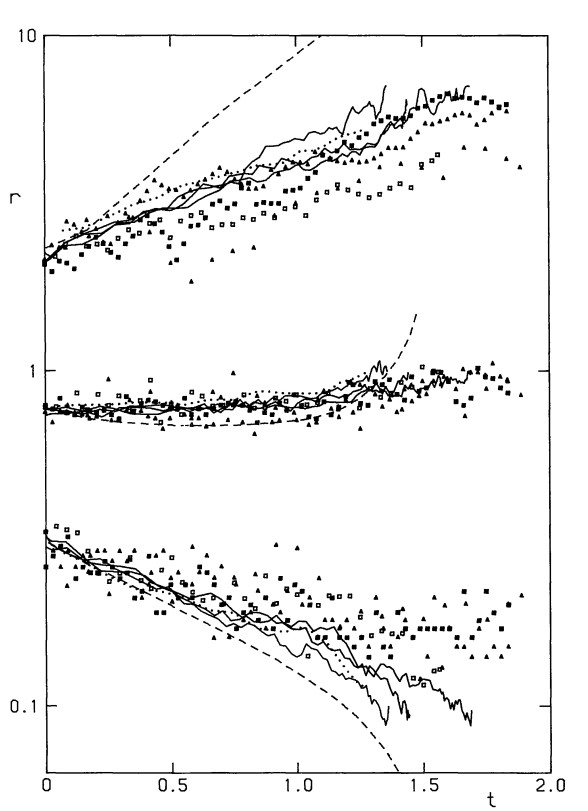}
 \caption{ Radii containing 10\%, 50\%, 90\% of the mass, plotted versus time for a cluster with stars of equal masses. Open triangles and squares: \nbody{} integration with $N=100$ and $N=250$ (Wielen). Filled triangles and squares:  \nbody{} Integration with $N=250$ (Aarseth). Full lines: Monte Carlo models (H{\'e}non1971). Dotted lines: Monte Carlo model (Shull and Spitzer). Dashed lines: fluid-dynamical model (Larson). From: \cite{Aarsethetal1974}; for citations and further discussion see that paper.} 
\label{fig:aarseth1974}
\end{figure}

Another line of research, through which the author of this article got the initial opportunity to collaborate with Sverre, started with a fluid dynamical model of star clusters, which was developed with my then supervisor Erich Bettwieser \citep{BettwieserSpurzem1986,LouisSpurzem1991}. The model was based on earlier ideas by \cite{Larson1970a,Larson1970b}, and it led to the detection of gravothermal oscillations \citep{SugimotoBettwieser1983,BettwieserSugimoto1984}. There was doubt about the validity of such models for star cluster evolution, a collaboration with Sverre started to verify the models against direct \nbody{} simulations. Sverre had already published a pioneering paper about comparison between \nbody{}, Monte Carlo, and fluid dynamical models \citep{Aarsethetal1974}; in that comparison though the fluid dynamical models did not give a very good agreement with \nbody{} 
(cf. Fig.~\ref{fig:aarseth1974}). Further work by \cite{SpurzemAarseth1996,GierszHeggie1994a,GierszHeggie1994b,GierszSpurzem1994,GierszHeggie1996} established a very good agreement between predictions of statistical physics and direct \nbody{} simulations. \cite{Makino1996} could find gravothermal oscillations in \nbody{} using GRAPE.

\section{\nbody{} -- the growth}

\subsection{Hermite Scheme}
\label{Hermite}

The Hermite scheme and hierarchically blocked time-steps \citep{McMillan1986,MakinoAarseth1992} improved the performance on vector and parallel computers and turned out to be efficient for all of recent parallel and innovative hardware (general and special purpose parallel computers, GRAPE and GPU, see Sect.~\ref{parallel} below). The Hermite scheme uses the analytic \nbody{} expression for the Newtonian gravitational acceleration $\vec{a}_0$ and its time derivative $\vec{{\dot a}}_0$ at some time $t_0$, to extrapolate low-order predicted positions $\vec{x}_p$ and velocities $\vec{v}_p$ at the next time step $t>t_0$. Then again gravitational acceleration $\vec{a}_1$ and its time derivative $\vec{{\dot a}}_1$ at $t$ are computed, using only predicted data. 
These four values of $\vec{a}$ and $\vec{{\dot a}}$ on two time points $t_0,t$ allow the the Hermite extrapolation to  compute the \nth{2} and \nth{3} derivatives of acceleration. These are used finally to obtain the high order corrected new positions and velocities from
\begin{align}
 \vec{x}(t) &= \vec{x}_p(t) + \frac{1}{24}(t-t_0)^4 \vec{a}_0^{(2)}
                                 +\frac{1}{120}(t-t_0)^5 \vec{a}^{(3)} + \co(\Delta t^6) \,\\
 \vec{v}(t) &= \vec{v}_p(t) + \frac{1}{6}(t-t_0)^3 \vec{a}_0^{(2)}
                                 +\frac{1}{24}(t-t_0)^4 \vec{a}_0^{(3)} + \co(\Delta t^5)\, .
\label{4.1.5}
\end{align}
The effective error in the methods is of order 
$\co (\Delta t^5)$, as was shown in test simulations \citep{Makino1991a}, which is not obvious due to the intrinsic low order predicted values used in the Hermite step.

\subsection{Time-step choice}

\cite{Aarseth1985a} provides an empirical time-step criterion
\begin{equation}
\Delta t = \sqrt{\eta \frac{ \vert\vec{a}\vert \vert\vec{a}^{(2)}\vert
                          + \vert\vec{{\dot a}}\vert^2 }{
                          \vert\vec{{\dot a}}\vert \vert\vec{a}^{(3)}\vert
                          + \vert\vec{a}^{(2)}\vert^2 }} \ .
\label{4.1.6}
\end{equation}
The error is governed by the choice
of $\eta$, which in most practical applications is taken to be
$\eta = 0.01 - 0.04$.
In \cite{Makino1991a} a different time
step criterion has been suggested, which appears simpler and more
straightforwardly defined, and couples the time-step to the difference
between predicted and corrected coordinates.
The standard Aarseth time-step criterion from Eq.~\eqref{4.1.6} has been used in
most $N$-body simulations so far, because it seems to achieve an optimal step better than (on average) the mathematically more sound Makino step (see the time-step related discussion in \citealt{Sweatman1994}). For the purpose of vectorization and parallelization the time steps of particles are quantized, only a variation by a factor of 2 is permitted, and the time $t$ should be an integer multiple of all time steps \citep{McMillan1986}.

\subsection{Ahmad-Cohen neighbour scheme}
For every particle a neighbour radius is defined, accelerations and their time derivatives according to the Hermite scheme are separately computed for particles inside and outside the neighbour radius. Consequently two different time steps are defined, the smaller irregular one related to the more fluctuating neighbour particle's forces, and another more steady, regular one related to the distant particles. The Ahmad-Cohen (AC) neighbour scheme goes back to \cite{AarsethZare1974} and typically yields regular timesteps, which are 5-10 times larger than the irregular steps. During irregular steps the regular forces are extrapolated using low order Taylor series. This reduces the amount of full force calculations significantly \cite[see][and also an earlier analysis by \citealp{MakinoHut1988}]{Huangetal2016}. The AC scheme is currently rarely used elsewhere, with the exception of a recent simulation of star clusters in a cosmological context using Enzo-N \citep{Joetal2024}. Note that the AC scheme is algorithmically similar to smoothed particle hydrodynamics \cite[see for a more historical quote][]{Spurzemetal2007}. Note also that the concept of a neighbour radius is not strictly implemented, there are precautions to search for fast incoming particles in a larger environment already in \nbody{5}, and nowadays a mass ratio far away from unity (in both directions) triggers larger search radii for neighbours in the current AC scheme \citep{Wangetal2015,Kamlahetal2022a,SpurzemKamlah2023}.

\begin{figure}
  \centering
  \includegraphics[width=0.7\textwidth]{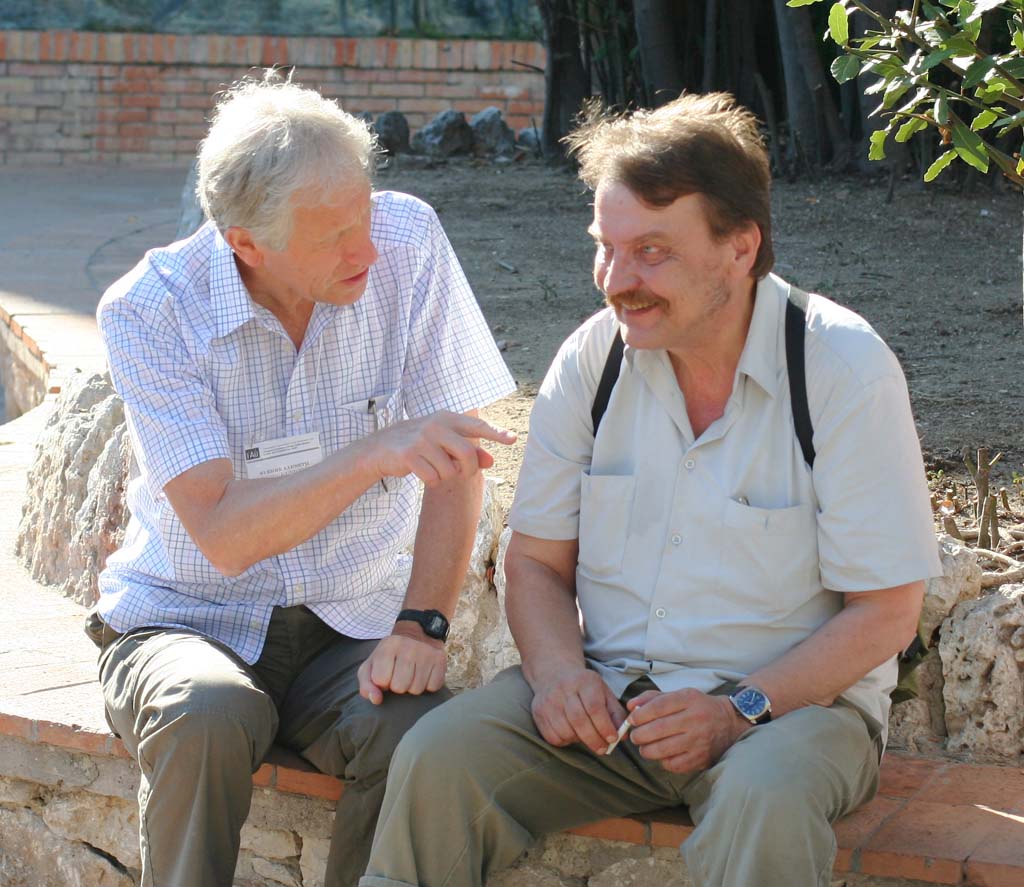}
 \caption{Sverre Aarseth and Seppo Mikkola, discussing chain regularization, in 2007.}
\label{fig:aarsethmikkola2007}
\end{figure}

\subsection{Regularizations}
\label{regularization}
As the relative distance $r$ of the two bodies becomes small, their time-steps are reduced to prohibitively small values, and
truncation errors grow due to the singularity in the gravitational potential \cite[][see Sect.~\ref{beginning}]{vHoerner1960,vHoerner1963,vHoerner2001}. The idea is to take both stars out of the main integration cycle, replace them by their centre of mass and advance the usual integration with this composite particle instead of resolving the two components. The internal motion of the two members of the regularized pair \cite[KS pair, for][]{KustaanheimoStiefel1965} is done in a new 4D coordinate system, which is similar to a quaternion space, including perturbing forces of other nearby stars. The theoretical base of this can be found e.g. in \cite{NeutschScherer1992}; the algorithm including the time transformation is in \cite{Mikkola1997a}. 

Close encounters between single particles and binary stars are
a central feature of cluster dynamics. The chain regularization \citep{MikkolaAarseth1998} is invoked if a KS pair has a  close encounter with another single star or another pair (see Fig.~\ref{fig:aarsethmikkola2007}). The chain regularization is effectively a chain of two body KS regularization, with the other particles perturbing the two body forces. 
Especially, if systems start with a large number of primordial (initial) binaries, such encounters may lead to stable (or quasi-stable) hierarchical triples, quadruples, and higher multiples.
They are treated by using special stability criteria \citep{MardlingAarseth2001}. Algorithmic regularization is a relatively recent different method based on a time transformed leap-frog method \citep{MikkolaMerritt2008}; it does not employ the KS transformation. See for its use and application the next subsection and Banerjee (this volume, on \nbody{7}).

\subsection{\nbody{} with Parallel and GPU computing}
\label{parallel}

The importance of vector and parallel computing was discussed in a pioneering conference \citep{McMillan1986}. 
The GRAPE chip \citep{Sugimotoetal1990,Makinoetal1993} is an application specific integrated circuit (ASIC), which computes in parallel gravitational forces between particles. It also computed the time derivative of the force, to be directly applicable to the Hermite scheme of \nbody{6} (\citealt{Makinoetal2003}).
A first implementation of the \nbody-code on GPU rather than GRAPE was reported by \cite{NitadoriAarseth2012} using CUDA kernels. Many of these kernels written by Keigo Nitadori are still in current use, even for the massively parallel programs such as \nbody{6++GPU}.

At the same time another development started, parallelization of \nbody{6} with the (at that time) new standard MPI (message passing interface), yielding \nbody{6++} \citep{Spurzem1999}.
\nbody{6++} parallelizes both force loops with MPI, for the regular and the neighbour force in the Ahmad--Cohen scheme.\\ 

With the advent of clusters, where nodes would be running MPI, and each node having a GPU accelerator, \nbody{6++GPU} was created -- on the top level MPI parallelization is done for the force loops (coarse grained parallelization) and at the bottom level each MPI process calls its own GPU to accelerate the regular force calculation \citep{Bercziketal2013}, and an AVX/SSE implementation on the CPUs accelerates prediction and neighbour (irregular) forces \citep{Wangetal2015}. With this code the \dragon{-I} simulation was done \citep{Wangetal2015,Wangetal2016}, at time of publication the award winning largest direct $N$-body simulation of a globular cluster with all required astrophysics (single and binary stellar evolution, stellar collisions, tidal field), simulated over 12 Gyrs.

In recent years, inspired also by LIGO/Virgo/KAGRA gravitational-wave detections \citep{Abbottetal2016a,Abbott2024}, numerous current updates have been made with regard to stellar evolution of massive stars in singles and binaries \cite[summarized in][]{Kamlahetal2022b,SpurzemKamlah2023}, and with regard to collisional build up of stars (mass loss at stellar collisions allowed) and intermediate-mass black holes \citep{Rizzutoetal2021,Rizzutoetal2022,ArcaSeddaetal2021a}.
The \dragon{-II} simulations using \nbody{6++GPU} show nicely that the ensemble of binary black hole mergers observed by LIGO/Virgo/KAGRA (including the most massive ones) can be well reproduced from star cluster simulations \citep{ArcaSeddaetal2023,ArcaSeddaetal2024a,ArcaSeddaetal2024b}. \dragon{-III} simulations are in progress for globular and nuclear star clusters \citep[][Cho et al., this volume]{Wuetal2025}. Recent work has brought the largest published number of particles to $1.8\cdot 10^6$ \citep{BarberAntonini2025,Rantalaetal2025}, using \petar and \bifrost codes, respectively (see below Sect.~\ref{newapproaches}). Current construction sites regarding \nbody{6++GPU} are a better model for neutron stars (Song et al. this volume), and for tidal disruption events near massive black holes \cite[Cho et al. this volume, and][]{Leeetal2025}. Finally a variant called \nbody{6++GPU-massless} has been developed which gives the possibility to integrate large numbers of very small mass particles (free floating planets, asteroids, or comets) inside a star cluster \citep{Flamminietal2025a,Flamminietal2025b}.
The current standard code \nbody{6++GPU} is available via GitHub\footnote{\url{https://github.com/nbody6ppgpu}}.

\begin{table}[ht!]
\begin{center}
\caption{Features of \nbody{1-7}, adopted with some updates from \cite{SpurzemKamlah2023}. }
\let\V\Checkmark
    \begin{tabular}{|p{2.5cm}|c|c|c|c|c|c|c|c|c|} \hline
                        & ITS & ACS & KS & HITS & PN & AR & CC & MPI & GPU \\\hline
    \nbody{1}     & \V  &     &    &      &    &    &    &     &    \\
    \nbody{2}     &     & \V  &    & \V   &    &    &    &     &    \\
    \nbody{3}     & \V  &     & \V &      &    &    &    &     &    \\
    \nbody{4}     &     &     & \V & \V   &    &    & \V &     &    \\ 
    \nbody{5}     & \V  & \V  & \V &      &(\V)&    & \V &     &    \\
    \nbody{6}     &     & \V  & \V & \V   &    &    & \V &     &    \\
    \nbody{6GPU}  &     & \V  & \V & \V   & \V &    & \V &     & \V \\
    \nbody{6++}   &     & \V  & \V & \V   &    &    & \V & \V  &    \\
    \nbody{6++GPU}&     & \V  & \V & \V   & (r)&    & \V & \V  & \V \\
    \nbody{7}     &     & \V  & \V & \V   & \V & \V &    &     & \V \\ \hline
    \end{tabular}\\
\V:  Included in standard version of that level \\
(\V): only included in special version of the code \citep{Kupietal2006}\\
(r): restricted, only orbit averaged 2.5PN \citep{ArcaSeddaetal2021a}\\
     \begin{tabular}{|p{0.7cm}|l|} \hline
ITS & Individual time-steps \citep{Aarseth1985a} \\
ACS & Ahmad--Cohen neighbour scheme \citep{AhmadCohen1973}\\
KS & KS--regularization of few-body subsystems \citep{KustaanheimoStiefel1965} \\
HITS & Hermite scheme and block steps \citep{MakinoAarseth1992}\\
PN & Post-Newtonian \citep{Kupietal2006,MikkolaMerritt2008,Aarseth2012}\\
AR & Algorithmic regularization \citep{MikkolaMerritt2008} \\
CC & Classical chain regularization \citep{MikkolaAarseth1998} \\
MPI & Message Passing Interface, multi-node multi-CPU parallelization \citep{Spurzem1999} \\
GPU & use of GPU acceleration \citep{NitadoriAarseth2012,Bercziketal2013}  \\ \hline
\end{tabular}
\label{versions}
 \let\V\undefined
  \end{center}
\end{table}

\section{New approaches}
\label{newapproaches}

There are two major new developments, one being \petar, the other \bifrost. \petar \citep{WangIwa2020} is a hybrid $N$-body code, which combines the P${}^3$T (particle-particle particle-tree) method and a slow-down time transformed symplectic integrator \citep[\textsc{SDAR}][]{WangNita2020}. The code is conceptually ahead of \nbody{6++GPU} in several respects; parallelization of a large number of hard binaries is included and a domain decomposition makes it easier to go to particle numbers much larger than $10^6$, as well as the use of the \tree scheme for distant groups of particles, rather than the Ahmad--Cohen neighbour scheme in \nbody{6++GPU}. But a detailed mutual verification and performance comparison of \petar and 
\nbody{6++GPU} for a large scale and long simulation including all astrophysical effects (such as e.g. the \dragon{-I} simulations) is still missing. Another novel approach is based on forward symplectic integrators \cite[FSI, cf. e.g.][]{DehnenHernandez2017}. It is called \bifrost \citep{Rantalaetal2021,Rantalaetal2023}, uses MPI parallelization and GPU acceleration and shows as well competitive benchmarks for very large $N$ (million and more). Both codes have been successfully used for large direct \nbody{} simulations \citep{Liuetal2024,BarberAntonini2025,Rantalaetal2024,Rantalaetal2025}.

\goodbreak\bigskip

\begin{acknowledgements}
Most importantly, the author wants to express thanks and gratitude to Sverre Aarseth for teaching, collaboration, and friendship over many years. Be assured Sverre, that our community will follow your friendly and open-minded spirit and keep your legacy alive. Furthermore I want to
thank Kai Wu, Francesco Flammini Dotti, Manuel Arca Sedda, Jarrod Hurley, Abbas Askar, Sambaran Banerjee, Peter Berczik, Roberto Capuzzo-Dolcetta, Mirek Giersz, Shuo Li, Shiyan Zhong, Xiaoying Pang, Thorsten Naab, and several more of co-authors not mentioned here for collaboration and helpful comments. Support by German Science Foundation (DFG, Sp 345/24-1), National Science Foundation of China (NSFC, grant No. 12473017), and Chinese Academy of Sciences (CAS) NAOC International Cooperation Office for its support in 2023, 2024, and 2025 is acknowledged. This research was supported in part by grant NSF PHY-2309135 to the Kavli Institute for Theoretical Physics (KITP).), and the Gauss Centre for Supercomputing e.V. (\url{https://www.gauss-centre.eu/}) for providing computing time through the John von Neumann Institute for Computing (NIC) on the GCS Supercomputer JUWELS and JUWELS-Booster at the  J\"ulich Supercomputing Centre in Germany (JSC). 
 The GRAPE team at Tokyo University, represented first by Daiichiro Sugimoto and later by Junichiro Makino deserves sincere acknowledgment for making GRAPE hardware available early for international collaborators.
\end{acknowledgements}

\bibliographystyle{iaulike}
\bibliography{refs.bib}

\end{document}